# Reconstructing the Nonlinear Filter Function of LILI-128 Stream Cipher Based on Complexity

Xiangao Huang[1]   Wei Huang[2]   Xiaozhou Liu[3]
Chao Wang[4]   Zhu jing Wang[5]   Tao Wang[1]

[1] College of Engineering, Shantou University, Shantou, 515063, China
Email: xghuang@stu.edu.cn
[2] Xian Electron Technique Research Institute of ZTE, 710061, China
[3] School of Electronics and Information Engineering, Xi' an Jiaotong University, Xi'an 710049, China
[4] College of Information, Chang an University, 710064, China
[5] Chongqing Communication institute, chongqing, 400035, China

*Abstract*: In this letter we assert that we have reconstructed the nonlinear filter function of LILI-128 stream cipher on IBM notebook PC using MATLAB. Our reconstruction need approximately $2^{12} \sim 2^{13}$ bits and the attack consumes 5825.016 sec (using tic and toc sentences of MATLAB) or 5825.016/3600=1.6181hours. We got the expression of the nonlinear filter function $f_d$ of Lili-128 which has 46 items from liner items to nonlinear items based on complexity, the phase space reconstruction, Clustering and nonlinear prediction. We have verified our reconstruction result correctness by simulating the overview of Lili-128 keystream generators using our getting $f_d$ and implement designers' reference module of the Lili-128 stream cipher, and two methods produce the same synchronous keystream sequence on same initial state, so that our research work proves that the nonlinear filter function of LILI-128 stream cipher is successfully reconstructed.

## 1. Introduction

LILI-128 stream cipher is designed by Dawson, Clark, Golic, Millan, Penna and Simpson[1], and submitted to NESSIE as a candidate cipher. Because the keystream sequence are a long period around $2^{128}$, high linear complexity which is conjectured to be at least $2^{68}$, and good statistics regarding the distribution of zeroes and ones, so designers assert that the LILI-128 keystream generator can resist currently known styles of attack. Some methods[2-5] of breaking it have been proposed, since LILI-128 stream cipher was publicized in 2000. The methods have already been shown that attack the LILI-128 stream cipher more efficiently than an exhaustive search for its secret key. However most of the attack methods consider only the complexity of time or memory for search for its secret key. For example, Time-Memory Tradeoff Attack[2] needs approximately $2^{46}$ bits, and Correlation Attack[3] needs approximately about $2^{23}$ bits. While Algebraic Attack [4] needs approximately $2^{18}$ bits. Even a new attack method[5] requires a mere $2^{7}$ bits of keystream. But this attack methods needs large memory and computations. For example, the paper[5] needs $2^{99.1}$ computations. In 2005, designers summarize all published styles of attack on the LILI-II stream cipher, and assert that LILI-II remains unbroken[6]. They encourage further analysis of the LILI-II stream cipher.

Section Ⅱ of this paper describes the structure of LILI-128. Section Ⅲ give the expression of the nonlinear filter function $f_d$. Section Ⅳ verifies the correctness of the expression of the nonlinear filter function $f_d$. We give the conclusion of this paper in Section Ⅴ.

## Ⅱ. The structure of LILI-128

The structure of the LILI-128 keystream generators is illustrated in Figure 1. The generator can be divided into two subsystems based on the functions they perform: the clock control subsystem and data generation subsystem. The clock control subsystem produces an integer sequence that is used to control the clocking of the second subsystem. The feedback polynomial of the LFSR$_c$ is chosen to be the primitive polynomial

$$G_c(x) = x^{39} + x^{35} + x^{33} + x^{31} + x^{17} + x^{15} + x^{14} + x^2 + 1 \quad (1)$$

Since $G_c(x)$ is primitive, the LFSR$_c$ produces a maximum-length sequence of period $P_c = 2^{39} - 1$. The function $f_c$ takes two bits as input and produced an integer $c_k$, such that $c_k \in \{1,2,3,4\}$. The value of $c_k$ is calculated as

$$c_k = f_c(y_1, y_2) = 2y_1 + y_2 + 1, \qquad k \geq 1. \quad (2)$$

The LFSR$_d$ is clocked by $c_k$ at least once and at most four times between the output of consecutive bits. The feedback polynomial of LFSR$_d$ is given as follows:

$$G_d(x) = x^{89} + x^{83} + x^{80} + x^{55} + x^{53} + x^{42} + x^{39} + x + 1 \quad (3)$$

Since $G_d(x)$ is a primitive polynomial, a period of $2^{89} - 1$ at maximum is guaranteed for LFSR$_d$ output sequence. The contents of 10 different stages of LFSR$_d$ are input to a nonlinear filter function $f_d$. The output of $f_d$ is the keystream sequence.

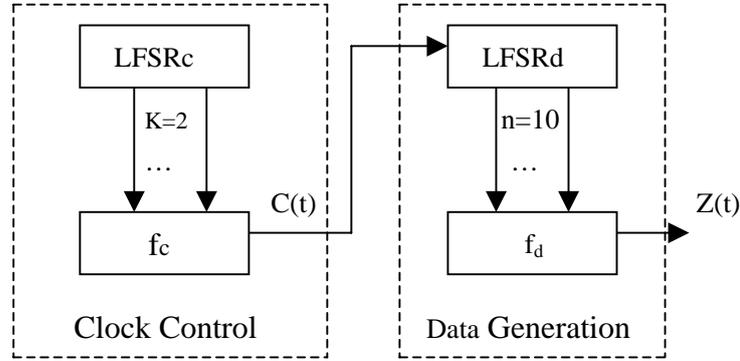

Figure 1　Overview of LILI-128 keystream generators

## Ⅲ. The expression of the nonlinear filter function $f_d$

　　The designers publicize all the structure of the clock control subsystem and the most part structure of the data generation subsystem. They do not publicize the expression of the nonlinear filter function $f_d$ in the data generation subsystem. The aim which we attack LILI-128 is the expression of the nonlinear filter function $f_d$. Up to now, all attacks explain only that LILI-128 has lower complexity as designers claim, and do not get the expression of the nonlinear filter function $f_d$. We get a reference module of LILI-128 from the Information Security Institute's webpage: http://www.isi.qut.edu.au/resources/lili/. We implement the reference module by using ASCII initial value "yyyyyyyyyyyyyyyy". First, we determine the least keystream bit amount of expressing the nonlinear filter function $f_d$ from the LILI-128 kystream sequence by means of measuring the complexity of the LILI-128 kystream sequence. The least key stream bit amount of expressing the nonlinear filter function $f_d$ is not equal for differ initial state. The least keystream bit amount is approximately $2^{12} \sim 2^{13}$ bits. We reconstruct the phase space diagram of the 10 dimensions: (1,2,3,4,5,6,7,8,9,10) from least keystream bit amount, and got the nonlinear filter function $f_d$ of Lili-128 which has 46 items from liner items to nonlinear items by clustering and nonlinear prediction under ASCII initial value "yyyyyyyyyyyyyyyy". The nonlinear filter function $f_d$ is expressed as follows:

$f_d = x^5 + x^4 + x^3 + x^2 + x^{10} * x^6 + x^{10} * x^4 + x^9 * x^3 + x^9 * x^1 + x^8 * x^2 + x^8 * x^1 + x^7 * x^6 + x^{10} * x^9 * x^5 + x^{10} * x^9 * x^4 + x^{10} * x^9 * x^3 + x^{10} * x^9 * x^2 + x^{10} * x^8 * x^4 + x^{10} * x^8 * x^3 + x^{10}$

$* x^7 * x^6 + x^{10} * x^7 * x^5 + x^{10} * x^7 * x^4 + x^9 * x^8 * x^6 + x^9 * x^8 * x^3 + x^9 * x^7 * x^6 + x^9 * x^7 * x^4 + x^9 * x^7 * x^3 + x^{10} * x^9 * x^8 * x^6 + x^{10} * x^9 * x^8 * x^4 + x^{10} * x^9 * x^8 * x^3 + x^{10} * x^9 * x^8 * x^1 + x^{10} * x^9 * x^7 * x^6 + x^{10} * x^9 * x^7 * x^4 + x^{10} * x^9 * x^7 * x^2 + x^{10} * x^8 * x^7 * x^5 + x^{10} * x^8 * x^7 * x^3 + x^9 * x^8 * x^7 * x^4 + x^9 * x^8 * x^7 * x^2 + x^9 * x^7 * x^6 * x^5 + x^9 * x^7 * x^6 * x^4 + x^{10} * x^9 * x^8 * x^7 * x^4 + x^{10} * x^9 * x^8 * x^7 * x^3 + x^{10} * x^9 * x^7 * x^6 * x^5 + x^{10} * x^9 * x^7 * x^6 * x^4 + x^9 * x^8 * x^7 * x^6 * x^5 + x^9 * x^8 * x^7 * x^6 * x^4 + x^{10} * x^9 * x^8 * x^7 * x^6 * x^5 + x^{10} * x^9 * x^8 * x^7 * x^6 * x^4$

(4)

The 10 inputs to $f_d$ are taken from LFSR$_d$ positions according to this full positive difference set: (0,1,3,7,12,20,30,44,65,80)[1] in the LILI-128, so we can map the set:(1,2,3,4,5,6,7,8,9,10) into the set:(0,1,3,7,12,20,30,44,65,80), and get another expression of the nonlinear filter function $f_d$.

## IV. Verifying the correctness for the expression of the nonlinear filter function $f_d$

If we describe LILI-128 keystream generator by MATLAB, the stages of LFSR$_c$ be labeled s[1], s[2],…,s[39] from left to right. At every time, we have the following formula to calculate the feedback bit:

w=s[38] ⊕ s[26] ⊕ s[25] ⊕ s[23] ⊕ s[9] ⊕ s[7] ⊕ s[5] ⊕ s[1]

where ⊕ indicates the addition modulo 2. Let the LFSR$_c$ shift left, and s[38]=w. Sequentially circulating and calculating, the LFSR$_c$ will produce a linear pseudorandom sequence. The function $f_c$ is given by

(5)

The stages of LFSR$_d$ be labeled u[1], u[2],…,u[89] from left to right. At time t, the feedback bit is calculated by the following formula

w=u[89] ⊕ u[51] ⊕ u[48] ⊕ u[37] ⊕ u[35] ⊕ u[10] ⊕ u[7] ⊕ s[1]

where ⊕ indicates the addition modulo 2. Let the LFSR$_d$ shift left, and u[89]=w. According to above circulation, the LFSR$_{cd}$ will produce a linear pseudorandom sequence. We map the set: (1,2,3,4,5,6,7,8,9,10) into the set: (1,2,4,8,13,21,31,45,66,81), the $f_d$ is given by:

$f_d = x^{13} + x^8 + x^4 + x^2 + x^{81} * x^{21} + x^{81} * x^8 + x^{66} * x^4 + x^{66} * x^1 + x^{45} * x^2 + x^{45} * x^1 + x^{31} * x^{21} + x^{81} * x^{66} * x^{13} + x^{81} * x^{66} * x^8 + x^{81} * x^{66} * x^4 + x^{81} * x^{66} * x^2 + x^{81} * x^{45} * x^8 + x^{81} * x^{45} * x^4 + x^{81} * x^{31} * x^{21} + x^{81} * x^{31} * x^{13} + x^{81} * x^{31} * x^8 + x^{66} * x^{45} * x^{21} + x^{66} * x^{45} * x^4 + x^{66} * x^{31} * x^{21} + x^{66} * x^{31} * x^8 + x^{66} * x^{31} * x^4 + x^{81} * x^{66} * x^{45} * x^{21} + x^{81} * x^{66} * x^{45} * x^8 + x^{81} * x^{66} * x^{45} * x^4 + x^{81} * x^{66} * x^{45} * x^1 + x^{81} * x^{66} * x^{31} * x^{21} + x^{81} * x^{66} * x^{31} * x^8 + x^{81} * x^{66} * x^{31} * x^2 + x^{81} * x^{45} * x^{31} * x^{13} + x^{81} * x^{45} * x^{31} * x^4 + x^{66} * x^{45} * x^{31} * x^8 + x^{66} * x^{45} * x^{31} * x^2 + x^{66} * x^{31} * x^{21} * x^{13} + x^{66} * x^{31} * x^{21} * x^8 + x^{81} * x^{66} * x^{45} * x^{31} * x^8 + x^{81} * x^{66} * x^{45} * x^{31} * x^4 + x^{81} * x^{66} * x^{31} * x^{21} * x^{13} + x^{81} * x^{66} * x^{31} * x^{21} * x^8 + x^{66} * x^{45} * x^{31} * x^{21} * x^{13} + x^{66} * x^{45} * x^{31} * x^{21} * x^8 + x^{81} * x^{66} * x^{45} * x^{31} * x^{21} * x^{13} + x^{81} * x^{66} * x^{45} * x^{31} * x^{21}$

$$x^8 \tag{6}$$

We simulate the keystream sequence of LILI-128 using (6) under the initial values which are ASCII "yyyyyyyyyyyyyyyy" according to Figure 1. Compare simulating result with the result of implementing the reference module, and two methods produce the same synchronous keystream sequence on same initial state "yyyyyyyyyyyyyyyy". We continue to verify the nonlinear filter function $f_d$ again by initial state "gggggggggggggggg" and "123456789abcdefg", and obtain all the same synchronous keystream sequence, so that the verifying work indicates that we have got the nonlinear filter function of the LILI-128 stream cipher.

## Ⅴ. Conclusion

In the design of LILI-128, Designers made an attempt to confuse the linear pseudorandom binary sequence with the long period $P_c=2^{89}-1$ by the clock-control with the long period $P_c=2^{39}-1$ and get high linear complexity of keystream sequence through the nonlinear filter function $f_d$ with 46 items and the most algebraic order of 6 to withstand all kinds of currently known attack. The LILI-128 keystream generator certainly resists currently known styles of attack, but it does not withstand our attack. The currently known styles of attack based on time complexity and memory complexity of arithmetic. The attacks do not consider the complexity of keystream sequence oneself. We get the least bit amount of the attack by measuring the complexity of the keystream sequence of LILI-128 and the nonlinear filter function $f_d$ from the least keystream bit amount by the phase space reconstruction, Clustering and nonlinear prediction. We will research the secret key from keystream sequence of LILI-128. Our research work has made a great breakthrough in stream cipher analysis, and will bring the importance influence on stream cipher design. We only publish our research result and do not explain the specific theory and algorithm of attacking the nonlinear filter function of LILI-128 stream cipher. Reader can verify the $f_d$ in Eq.(4) first of all. And then $f_d$ is redesigned, the stream cipher sequence of LILI-128 output is obtained by simulating Figure 1 using a ASCII initial state. Reader can sends the ASCII initial state and the stream cipher sequence which length is $2^{13}$ bits to us. We will return $f_d$ to him.

**Acknowledgment**

The authors would like to thank Academician Changxiang Shen, Professor Yumin Wang, Guozhen Xiao and Dinyi Pei for their useful comments and suggestions.